\documentclass[reprint,aps,prl,twocolumn,superscriptaddress,amsmath,amssymb,floatfix]{revtex4-2}
\usepackage{graphicx}
\usepackage{layout}
\usepackage{natbib}
\usepackage{dcolumn}
\usepackage{braket}
\usepackage{bm}
\usepackage{color}
\usepackage{verbatim}
\usepackage{hyperref}

\begin{document}

\title{SWAP gate between a Majorana qubit and a parity-protected superconducting qubit}

\author{Luca Chirolli}
\affiliation{Department of Physics, University of California Berkeley, CA-94720, USA}
\affiliation{Istituto Nanoscienze - CNR, I-56127 Pisa, Italy}

\author{Norman Y. Yao}
\affiliation{Department of Physics, University of California Berkeley, CA-94720, USA}

\author{Joel E. Moore}
\affiliation{Department of Physics, University of California Berkeley, CA-94720, USA}
\affiliation{Materials Sciences Division, Lawrence Berkeley National Laboratory, Berkeley, California 94720, USA}

\begin{abstract}
High fidelity quantum information processing requires a combination of fast gates and long-lived quantum memories. In this work, we propose a hybrid architecture,  where a parity-protected superconducting qubit is directly coupled to a Majorana qubit, which plays the role of a quantum memory.  The superconducting qubit is based upon a $\pi$-periodic Josephson junction realized with gate-tunable semiconducting wires, where the tunneling of individual Cooper pairs is suppressed. One of the wires additionally contains four Majorana zero modes that define a qubit. We demonstrate that this  enables the implementation of a SWAP gate, allowing for the transduction of quantum information between the topological and conventional qubit. This architecture combines fast gates, which can be realized with the superconducting qubit, with a topologically protected Majorana memory.
\end{abstract}

\maketitle

{\it Introduction.---}
Majorana zero-energy modes (MZMs) \cite{majorana1937teoria,read2000paired,kitaev2001unpaired,ivanov2001non-abelian,wilczek2009majorana} realized in condensed matter systems ~\cite{lutchyn2010majorana,yuval2010helical,alicea2012new,beenakker2013search,albrecht2016exponential,aguado2017majorana,lutchyn2018majorana,prada2020from,flensberg2021engineered} are localized at far ends of the system and are expected to be robust with respect to local perturbations. Topologically protected Majorana qubits (MQs) encoded in MZM are hardly disturbed by an environment that acts locally and are expected to show very long coherence times \cite{kitaev2001unpaired}. Several schemes for topologically protected quantum information processing based upon MZMs have been proposed~\cite{bravyi2005universal,kitaev2003fault-tolerant,dassarma2005topological,freedman2006towards,nayak2008non-abelian,dassarma2015majorana,alicea2011non-abelian,aasen2016milestones}. An alternative route, which may be particularly relevant in the noisy intermediate-scale quantum technology era \cite{preskill2018quantum}, is to envision a hybrid architecture combining topologically protected memories with conventional qubits~\cite{nielsen_chuang_2010,kitaev2002classical,preskill1997fault-tolerant}.  In this context, a Majorana qubit could serve as a quantum memory or quantum buffer and a key ingredient is represented by a SWAP gate, that enables quantum state swapping between an ``ordinary'' qubit  and the memory.  
 
Superconducting qubits are a particularly natural candidate since solid state realizations of MZMs typically involve a parent superconductor and are often based on proximity-induced topological superconductivity. Early suggestions for interactions between a superconducting qubit and a MQ were based on a number of different schemes, including: 
coupling to a flux qubit \cite{jiang2011interface,bonderson2011topological}, a flux qubit based readout \cite{hassler2010anyonic,hou2011probing}, a top-transmon in which MZMs are moved between the islands of a transmon and a parity-protected scheme is discussed \cite{hassler2011top-transmon} and a fluxonium based on the $4\pi$-Josephson effect, where a MQ generates  avoided crossing in the flux-dependent spectrum \cite{pekker2013proposal}. Other approaches based on transmons have been recently put forward, where the fermion parity is encoded in the cavity dispersive shift \cite{ginossar2014microwave,yavilberg2015fermion}, and the microwave spectrum can be used to detect the presence of MZMs \cite{avila2020majorana,avila2020superconducting}. In this context, the role of Coulomb effects on an island hosting MZMs has also been recently considered \cite{pikulin2019coulomb}.

In this work, we study the coupling between a parity-protected superconducting qubit (PPSQ) and a Majorana qubit and we present a setup that enables the implementation of a SWAP gate between the two. A PPSQ is constituted by a superconducting island coupled to a reference circuit by a $\pi$-periodic Josephson junction (JJ) described by a $\cos(2\varphi)$  energy-phase relation. The latter captures the tunneling of pairs of Cooper pairs at the junction and preserves the parity of the Cooper pair number on the island. The Majorana qubit is formed through four MZMs, two of which contained in the superconducting island and the other two in the bulk reference and coupled at the junction is via the $4\pi$-periodic Josephson effect. In such a combined system, in the weak transmon regime the charging energy of the superconducting qubit can distinguish the presence of a single electron in the island, thus resolving the two states of the MQ. In addition, the coupling via the $4\pi$-periodic Josephson effect enables the implementation of a SWAP gate.

An experimental realization of an effective $\cos(2\varphi)$ relation has been a long-standing goal and only recently have parity-protected superconducting qubits  been realized. Relevant implementations are based on rhombi of four nominally equal JJs \cite{blatter2001design,doucot2002pairing,protopopov2004anomalous,gladchenko2009superconducting,bell2014protected}, ladders of JJs realizing current-mirror $0$-$\pi$ qubit \cite{kitaev2006protected}, gatemons realized through semiconducting wires \cite{delange2015realization,larsen2015semiconductor,luthi2018evolution,larsen2020parity-protected}, and superinductors obtained through chains of JJs \cite{brook2013protected,smith2020superconducting,gyenis2021experimental}. The use of $\cos(2\varphi)$ Josephson relations have recently been  proposed  as a tool for modulating the Josephson potential and enhancing  coherence times  through band engineering \cite{chirolli2021enhanced}. In the present work we focus on the gatemon-based realization, but the protocol is general and can be realized through any implementation of $\pi$-periodic JJs.

\begin{figure}[t]
\includegraphics[width=0.45\textwidth]{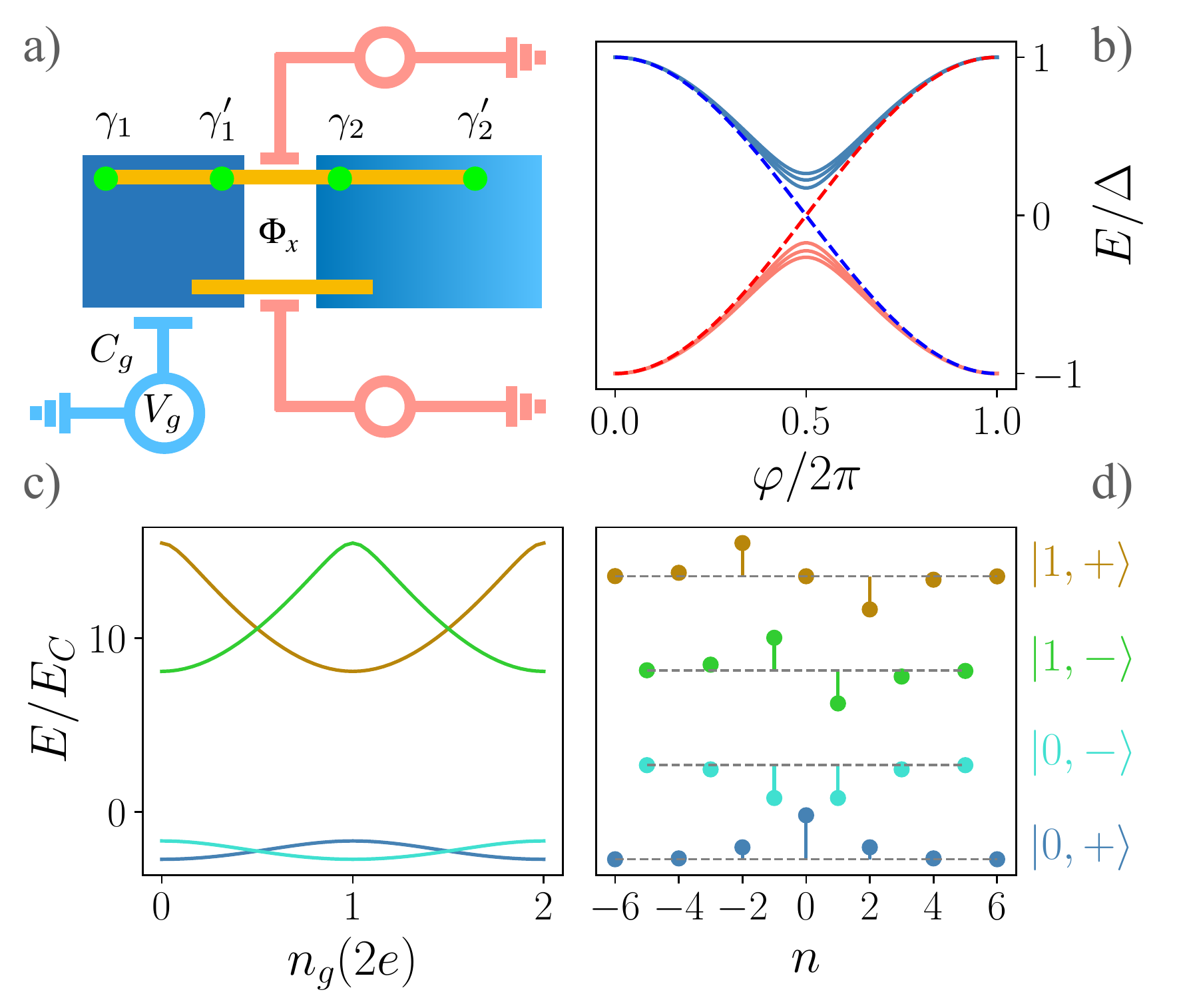}
\caption{a) Circuit of a Parity-Protected Transmon qubit coupled to a Majorana qubit.  b) Particle-hole symmetric subgap spectrum of a gate-tunable JJ containing a Majorana bound state and three Andreev bound states. c) Four lowest energy states $|k,\tau\rangle$ wave function, specified by principal quantum number $k$ and parity $\tau=\pm$, is shown in the charge basis showing the even and odd number content. d) Their energy versus offset charge $n_g=C_gV_g/(2e)$ for $E_J/E_C=10$. \label{Fig1}}
\end{figure}

{\it The system.---}
We start the analysis considering the system shown in Fig.~\ref{Fig1}a). It consists of a superconducting island coupled to a reference superconductor via two semiconducting wires that form two gate-tunable JJs  \cite{larsen2015semiconductor,luthi2018evolution}. One of the two wires is assumed to realize a topological superconductor \cite{lutchyn2010majorana,yuval2010helical} and it is cut in two segments, each hosting a pair of MZMs at its ends, $\gamma_i$ and $\gamma_i'$ for $i=1,2$.  The modes $\gamma_1'$ and $\gamma_2$ overlap across the junction and produce a bound state whose energy depends on the gauge invariant superconducting phase difference $\varphi$ via the $4\pi$-periodic Josephson effect, as shown by the dashed lines in Fig.~\ref{Fig1}b). In addition, the two segments are assumed to be of a length on order of the superconducting coherence length, so that $\gamma_i$ and $\gamma_i'$ in each segment hybridize and their energy splitting can be tuned by an in-plane Zeeman field, that induces an oscillatory behavior \cite{dassarma2012splitting,prada2012transport,rainis2013towards,albrecht2016exponential}.  The Hamiltonian describing the coupling among the four MZMs reads
\begin{equation}\label{Eq:MQ}
H_{\rm MZM}=i\gamma_1'\gamma_2 E_M\cos(\varphi/2)+i\lambda_1\gamma_1\gamma_1'+i\lambda_2\gamma_2\gamma_2', 
\end{equation}
with $E_M$ the bare coupling at the junction, and $\lambda_1$ and $\lambda_2$ the hybridization energy in each wire segment.  

In addition to the MZMs, the two semiconducting wires coupled to the superconducting island are assumed to host also highly transparent Andreev bound states, that in absence of an external magnetic field are described by an energy-phase relation ${\cal U}_i(\varphi_i)=-\Delta\sum_n\sqrt{1-T^{(i)}_n\sin^2(\varphi_i/2)}$, with $i=1,2$ labeling the two wires, $\Delta$ the induced superconducting gap on the wires, $T^{(i)}_n$ the transmission coefficient of the $n$-th conducting channel, and $\varphi_i$ the gauge-invariant phase difference across the $i$-th junction. The particle-hole symmetric subgap spectrum of the topological junction is schematically shown Fig.~\ref{Fig1}b), where the Andreev states (continuous lines) and the Majorana states (dashed lines) are shown together. In general, we assume the two Andreev states and the Majorana state to be independently tunable through applied electrostatic gates. A more precise tuning can be experimentally achieved through a parallel of three wires \cite{ronzani2014balanced}, one of which realizing a topological superconductor.

{\it Parity-protected superconducting qubit.---}
The parallel of two wires yields the effective Josephson energy-phase relation ${\cal U}(\varphi)={\cal U}_1(\varphi)+{\cal U}_2(\varphi_x-\varphi)$, with $\varphi_x=2\pi \Phi_x/\Phi_0$ the flux threading the loop in units of $\Phi_0=h/2e$.  It has been shown in Ref.~\cite{larsen2020parity-protected} that by tuning ${\cal U}_1={\cal U}_2$ and $\varphi_x=\pi$ all odd harmonics are suppressed by destructive interference and the junction realizes a $\pi$-periodic energy-phase relation \footnote{Strictly speaking, an exact correspondence between energy and transmission eigenvalues is lost in presence of an external magnetic field and a more complicate energy-phase relation is expected, with the appearance of odd terms such as $\sin(\varphi)$. }. Thus, the junction effectively realizes one $\pi$-periodic JJ. Focusing on the second harmonic and neglecting all other exponentially suppressed harmonics, the Hamiltonian of the PPSQ is specified by the Josephson energy and the charging energy and reads 
\begin{equation}\label{Eq:Hcos2phi}
H_0=4E_C(\hat{n}-n_g)^2-E_J\cos(2\hat{\varphi}).
\end{equation}
The gauge invariant phase difference $\hat{\varphi}$ and the number of Cooper pairs $\hat{n}$ are conjugate variables and satisfy canonical commutation rules $[\hat\varphi,\hat{n}]=i$. Here, $E_J$ is the Josephson energy of the $\pi$-periodic element and $E_C=e^2/2C$ is the charging energy, with the total capacitance, $C=C_J+C_g$, given by the capacitance of the JJ, $C_J$, and the capacitance $C_g$ of a side gate. The latter enables to control the average offset charge on the island $n_g=C_gV_g/(2e)$ through a voltage $V_g$.

In the charge basis, in which the charging term is diagonal, the $\pi$-periodic Josephson term is written as $\cos(2\hat{\varphi})=\frac{1}{2}\sum_n|n+2\rangle\langle n|+{\rm h.c.}$, and we clearly see that it does not couple states differing by an odd number of Cooper pairs in the superconducting island. We then define a conserved quantity, labeled by $\tau=\pm 1$, associated to the parity of the number of Cooper pairs in the island, that we term boson parity. The Hamiltonian of the PPSQ is written as $H_0=H^0_++H^0_-$, where
\begin{equation}\label{Eq:Hpm}
H^0_\pm=\sum_{n_\pm}4E_C(n_\pm-n_g)^2\mathbb{P}_{n_\pm}-\frac{E_J}{2}|n_\pm+2\rangle\langle n_\pm|+{\rm h.c.},
\end{equation}
with $n_\pm$ even/odd, respectively, $\mathbb{P}_{n_\pm}=|n_\pm\rangle\langle n_\pm|$.  The energy levels can be generally written as $E^\tau_k(n_g)$, with $k=0,1,2,\ldots$ a principal quantum number. Deep in the charging regime characterized by $E_C\gg E_J$, the system is very sensitive to the presence of additional charge on the island and few charge states are sufficient to describe the spectrum. The sensitivity persists also in the weak transmon regime, specified by $E_C/E_J\lesssim 1$, and the four lowest energy eigenvalues versus the offset charge and their associated wavefunctions are shown in Fig.~\ref{Fig1} c) and d), respectively, for $E_C/E_J=0.1$. By inspection of Eq.~(\ref{Eq:Hpm}) we see that a shift $n_g\to n_g+2$ leaves the Hamiltonian invariant, resulting in an overall $4e$ periodicity of the spectrum on the offset charge, as shown in Fig~\ref{Fig1}b). The crossings at $n_g=1/2,3/2$ are protected by conservation of the boson parity. Focusing on the four lowest energy states, the Hamiltonian $H_0$ then reads 
\begin{equation}\label{Eq:Hppsq}
H_0=\frac{\epsilon(n_g)}{2}\tau_z(1-\sigma_z)+\frac{1}{2}(E_{10}+\epsilon'(n_g)\tau_z)(1+\sigma_z),
\end{equation}
where the Pauli matrices $\sigma_i$ span the two states associated to $k=0,1$ and the Pauli matrix $\tau_i$ span the two boson parity sectors. In addition, we have $E_{10}=(E^+_1+E^-_1-E^+_0-E^-_0)/2$, $\epsilon=(E^-_0-E^+_0)/2$, $\epsilon'=(E^+_1-E^-_1)/2$. In Fig.~\ref{Fig1}c) deviations from a $\cos(\pi \hat{n}_g)$ dependence, typical of the transmon regime \cite{koch2007charge-insensitive}, can be appreciated in the higher energy levels, allowing for independent tuning of $\epsilon$ and $\epsilon'$ by the offset charge. 

The parity-protected qubit is provided by the two lowest energy states, corresponding to the choice $\sigma_z=-1$, and in order to separate them from the rest of the spectrum and  still keep sensitivity to the charge on the island an intermediate $E_C\lesssim E_J$ is required.

{\it Coupling Hamiltonian.---}
We now consider the coupling between the MQ and the PPSQ. The full Hamiltonian reads $H = H_0+H_{\rm MZM}$. To properly treat the MQ we introduce two fermionic operators, $c_i=(\gamma_i+i\gamma_i')/2$, that allow us to label states via their occupation number $|q_1,q_2\rangle$, with $\hat{q}_i=c^\dag_ic_i$. In each sector of the total fermion parity ${\cal P}=e^{i\pi(\hat{q}_1+\hat{q}_2)}$ the four MZMs encode a qubit degree of freedom \cite{nayak2008non-abelian}. We can introduce Pauli operators to span the qubit states, ${\cal P}\eta_x=i\gamma_1'\gamma_2$, $\eta_z=\hat{p}_1=-i\gamma_1\gamma_1'$, and ${\cal P}\eta_z=\hat{p}_2=-i\gamma_2\gamma_2'$.  The presence of MZMs in the superconducting island affects the spectrum of the system through a constraint in the wavefunction, that becomes anti-periodic in case an odd number of electrons is present in the island, $ \Psi(\varphi+2\pi n)=(-)^{n\hat{q}_1}\Psi(\varphi)$ \cite{fu2010electron}. To make manifest the boundary condition we perform a unitary transformation $H\to U^\dag H U$, with $U=e^{i\varphi\hat{q}_1/2}$ \cite{vanheck2011coulomb,ginossar2014microwave,keselman2019spectral,avila2020superconducting}, and Hamiltonian becomes
\begin{eqnarray}
    H &=& 4E_C(\hat{n}+\hat{q_1}/2-n_g)^2-E_J\cos(2\hat{\varphi})-\lambda_1\hat{p}_1-\lambda_2\hat{p}_2\nonumber\\
    &+&E_M\left[(c_1c_2 -c_2^\dag c_1)(1+e^{i\hat{\varphi}/2})+{\rm h.c.}\right].
\end{eqnarray}
The coupling between the MQ and the PPSQ appears now in two terms: i) the charging energy and ii) the hybridization term proportional to $E_M$. In each fermion parity sector the charge $\hat{q}_1$ takes the values $0,1$ between the MQ states and the hybridization energy only differs for the sign of the coupling in the two fermion parity sectors.  

\begin{figure}[t]
\includegraphics[width=0.45\textwidth]{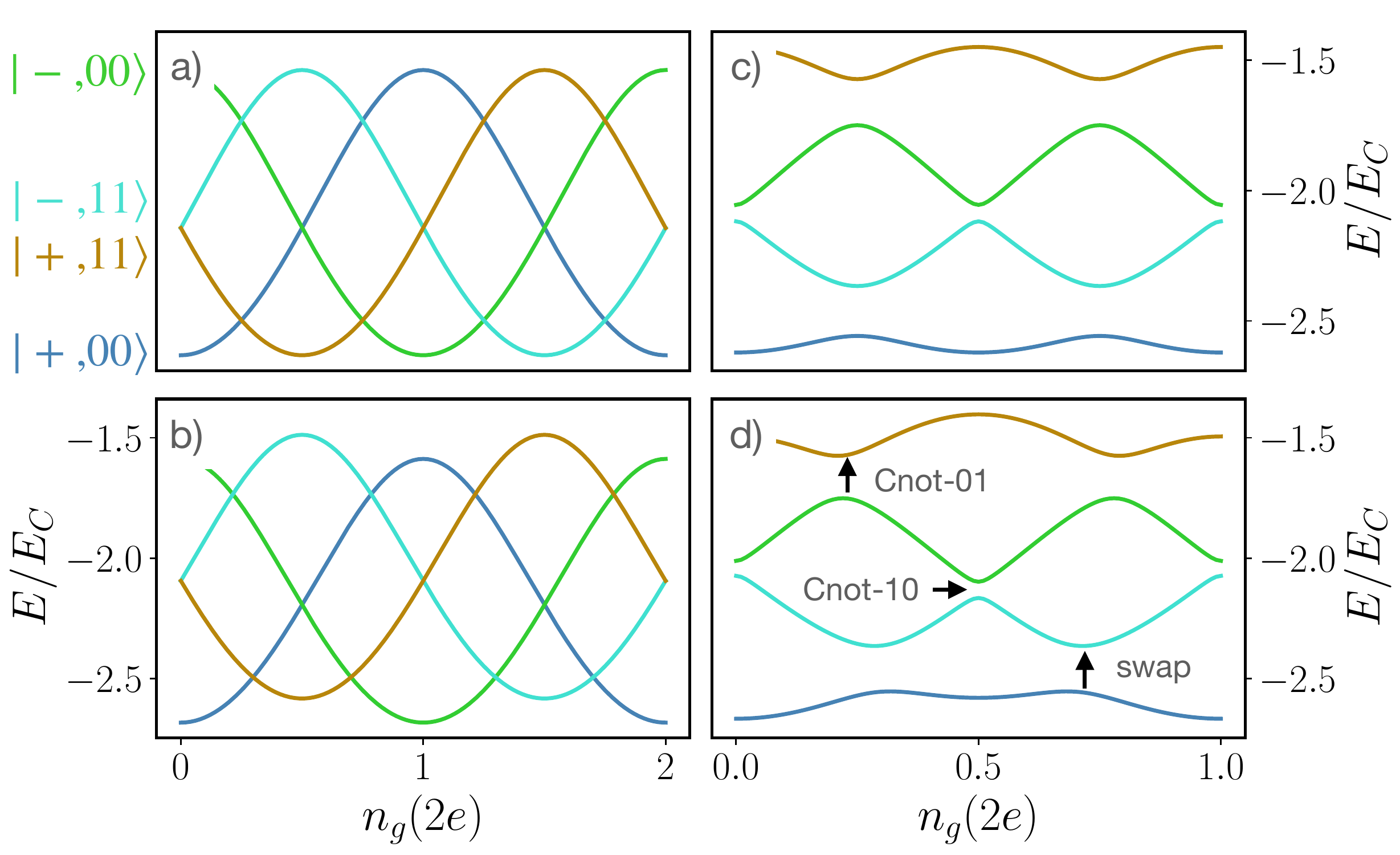}
\caption{Four lowest energy levels of the Hamiltonian Eq.~(\ref{Eq:Hmat}) shown as a function of the offset charge $n_g$ for equal JJs, $\varphi_x=\pi$, and  an effective $E_J/E_C\simeq 10$.  a) Unperturbed case $\lambda=E_M=0$. b) $\lambda = 0.2 E_C$ and $E_M=0$.  c) A finite $E_M=0.2 E_C$ splits the crossings at $n_g=0,1/4,1/2$ and results in a $e$-periodic spectrum. d) A finite $\lambda$ shifts the avoided crossing from $n_g=1/4,3/4$ and results in a $2e$-periodic spectrum. The three avoided crossings allow for two C-not gates and a SWAP gate, as indicated. \label{Fig2}}
\end{figure}

Without loss of generality we consider the even fermion parity sector (${\cal P}=1$). By taking matrix elements between the two states $|00\rangle, |11\rangle$ of the MQ the Hamiltonian reads
\begin{equation}\label{Eq:Hmat}
H=\left(\begin{array}{cc}
H_0(n_g)-\lambda & E_M(1+ e^{i\hat{\varphi}})/2\\
E_M(1+e^{-i\hat{\varphi}})/2 & H_0(n_g-1/2)+\lambda
\end{array}\right).
\end{equation}
In the diagonal blocks, in addition to the shift in the offset charge, a further energy imbalance is provided by the hybridization energy of the MZMs in the same segment of the wire, $\lambda=\lambda_1+\lambda_2$. The four lowest energy states associated to the two bosonic parity states and the two MQ states are $|\psi^\pm_{n_g}\rangle|00\rangle$ and $|\psi^\pm_{n_g-1/2}\rangle|11\rangle$ and their energies as a function of $n_g$ are shown in Fig.~\ref{Fig2}a) for the choice $E_J\simeq 10 E_C$,  in the weak transmon regime  \footnote{Calculations are done with the parallel of JJs realizing equal Andreev states for the choice $T^{(i)}_1=0.96$, $T^{(i)}_2=0.91$, $T^{(i)}_3=0.9$, and $\Delta\simeq 72 E_C$.}. By switching on a finite $\lambda$ the $|11\rangle$ states shift up in energy with respect to the $|00\rangle$ states, as shown in Fig.~\ref{Fig1}b). The periodicity versus the offset charge is still $4e$, as no coupling is present between the different states. In Fig.~\ref{Fig2}c) a finite $E_M$ opens a splitting between all crossing producing a change in the periodicity from $4e$ to $e$ in the offset charge.  In Fig.~\ref{Fig2}d), by simultaneously switching on both $\lambda$ and $E_M$ we see that the periodicity of the spectrum becomes $2e$.

\begin{figure}[t]
\includegraphics[width=0.45\textwidth]{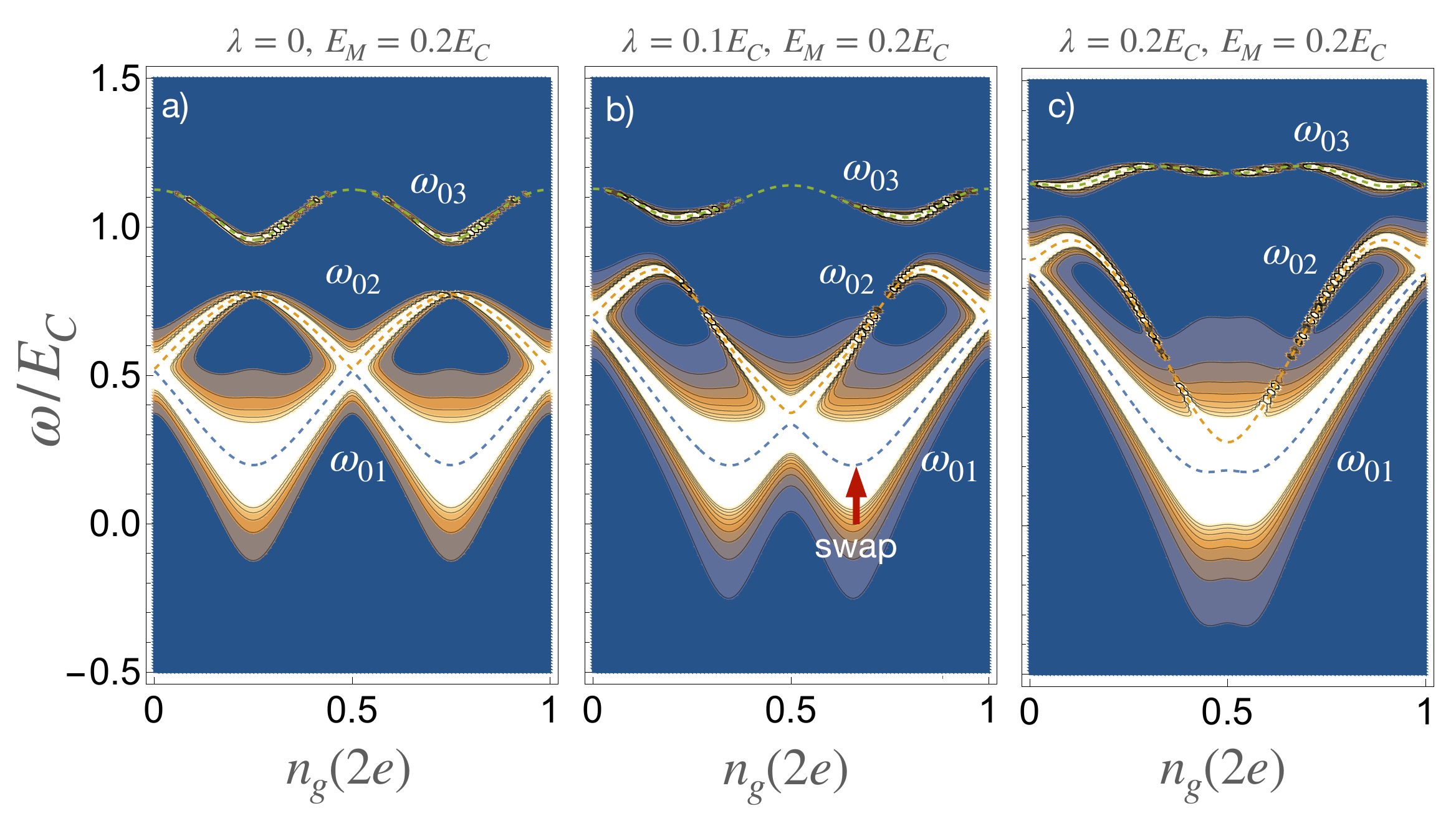}
\caption{Microwave absorption spectrum $S(\omega)$ of the coupled system versus the offset charge $n_g$ and the frequency $\omega$ in the weakly transmon for $E_J/E_C=10$ and $E_M=0.2 E_C$. a)  $\lambda=0$ yields $e$-periodic transitions. A finite  $\lambda=0.1$ b) and $\lambda=0.2 E_C$ c) produce a $2e$-periodic spectrum.    \label{Fig3}}
\end{figure}

The effect of the coupling parametrized by $E_M$ and $\lambda$ is best appreciated in the microwave spectrum. The latter measures the probability of photon absorption by a microwave cavity capacitively coupled to the superconducting island. The offset charge acquires a time-dependence $n_g\to n_g+\delta n_g(t)$ that gives rise to a coupling $\delta n_g(t)(\hat{n}+\hat{q}_1/2)$. Assuming the system initially in the ground state $|0\rangle$, the linear response microwave spectrum is given by $S(\omega)=\sum_{p>0}|\langle p|({\hat{n}+\hat{q}_1/2})|0\rangle|^2\delta(\omega-\omega_{0p})$, with $|p\rangle$ all eigenstates of the Hamiltonian Eq.~(\ref{Eq:Hmat}) and $\omega_{0p}=\omega_p-\omega_0$. In absence of MZMs the ground state switches boson parity with period $2e$. The microwave spectrum is shown in Fig.~\ref{Fig3} for $E_M=0.2 E_C$  and different values of $\lambda$.  In Fig.~\ref{Fig3}a) we see that a finite $E_M$ yields a $e$-periodic spectrum compatible with the avoided crossings between the energy levels shown in Fig.~\ref{Fig2}c). In Fig.~\ref{Fig3}b,c) a finite $\lambda$ breaks the $e$-periodicity and restores a $2e$-periodic spectrum compatible with the one shown in Fig.~\ref{Fig2}d).

{\it SWAP gate.---} 
We can write an effective low energy qubit-qubit Hamiltonian by taking matrix elements of the Hamiltonian Eq.~(\ref{Eq:Hmat}) between the four lowest energy levels, $|\psi^\pm_{n_g}\rangle|00\rangle$ and $|\psi^\pm_{n_g-1/2}\rangle|11\rangle$, of the Hamiltonian $H_0$. The relevant matrix elements induced by the MQ are $g_{\tau,\tau'}=\frac{E_M}{2}\langle\psi^\tau_{n_g}|(1+e^{i\hat{\varphi}})|\psi^{\tau'}_{n_g-1/2}\rangle$, and in the weak transmon regime have a weak dependence on $n_g$ that can be neglected. The Hamiltonian describing the coupling of the MQ with the PPSQ reads
\begin{equation}\label{Hswap}
H=\Omega(n_g)\tau_z+\lambda\eta_z+g(1+\tau_x)\eta_x+\delta(n_g)\eta_z\tau_z,
\end{equation}
where $g=g_{\tau,\tau'}$, $\Omega(n_g)=(\cos(\pi n_g)-\sin(\pi n_g))/2$, and $\delta(n_g)=(\cos(\pi n_g)+\sin(\pi n_g))/2$. This is the most important result of the present work. It shows that a coupling between a PPSQ and a MQ is possible and can be completely controlled by the hybridization terms of the MQ, $\lambda$ and $E_M$. In addition, it allows the realization of the two fundamental C-nots gates and the SWAP gate. The latter can be obtained by actively driving the flip-flop term $|+,11\rangle\langle-,00|$ at a frequency $\omega_{\rm swap}=2(\Omega-\lambda)$ or by tuning the offset charge in proximity of the avoided crossing indicated in Fig.~\ref{Fig2}d) and driving the adjacent microwave cavity capacitively coupled to the superconducting island at frequency $\omega \sim E_M$ it is possible to generates a SWAP gate, as confirmed by the microwave spectrum of Fig.~\ref{Fig3}b).

\begin{figure}[t]
\includegraphics[width=0.45\textwidth]{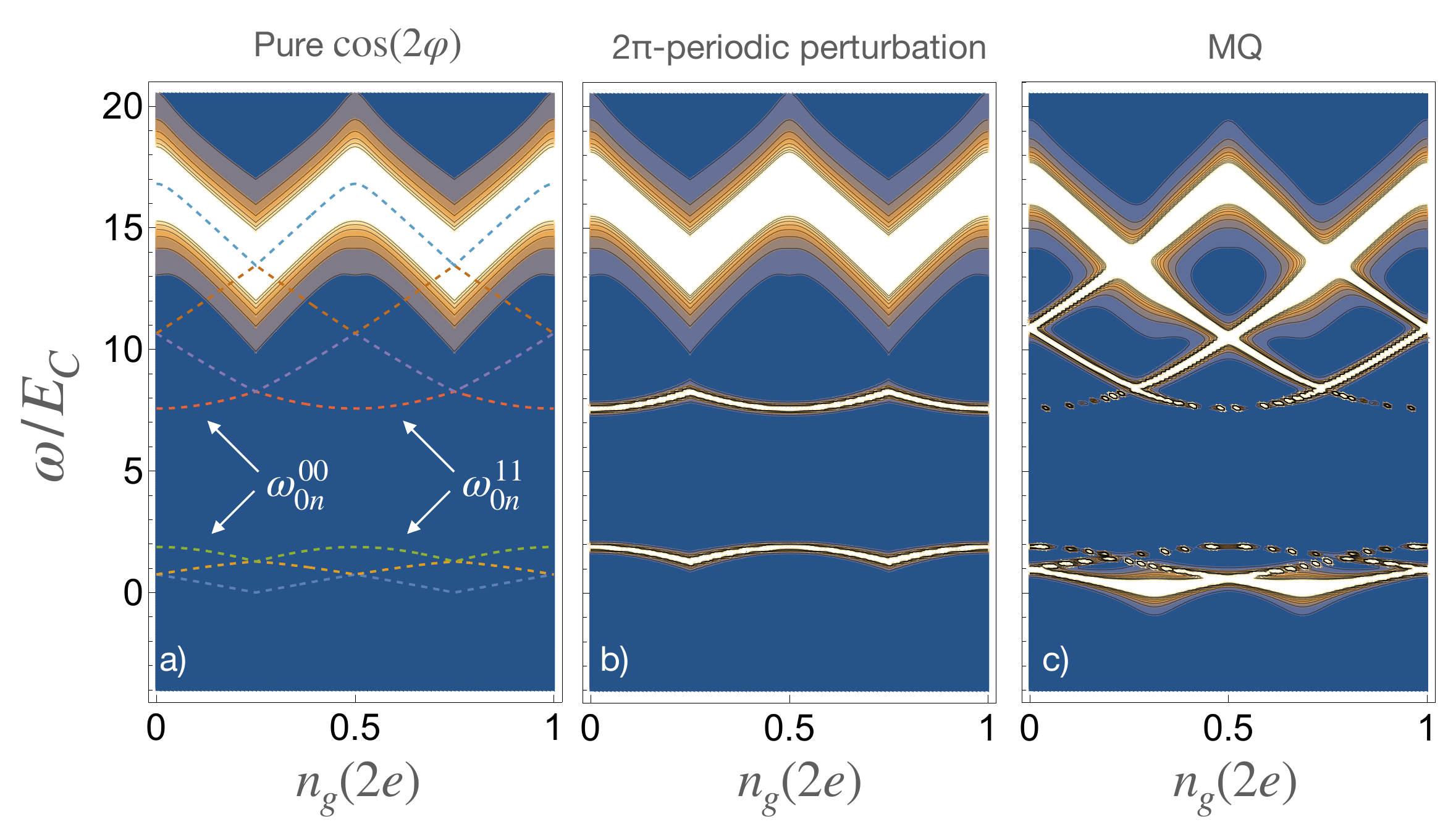}
\caption{Microwave spectrum associated to a) a pure $\pi$-periodic PPSQ, b) the effect of $2\pi$-periodic perturbations, and c) the effect of MZMs in one junction. The part of the spectrum between $-0.5<\omega/E_C<1.5$ corresponds to Fig.~\ref{Fig3}b). \label{Fig4}}
\end{figure}

{\it $2\pi$-periodic perturbations.---}
The ability to engineer a $\pi$-periodic JJ is crucial for the definition of a well behaved parity-protected qubit. Yet, in experimental realizations spurious $2\pi$-periodic perturbations are not negligible. It is then important to assess their impact. The minima of the $\cos(2\varphi)$ potential, located at $\varphi=0,\pi$, map into each other by the mirror transformation $M_\varphi:\varphi\to \pi-\varphi$. Even and odd perturbations under $M_\varphi$, such as  $\sin(\varphi)$ and $\cos(\varphi)$, respectively, yield different terms. An odd perturbation couples states of same principal quantum number $k$ and opposite $\tau$ yielding $\alpha_k = \langle k,-|\cos(\varphi)|k,+\rangle$, that produce an energy imbalance between the two minima.  Furthermore, states of opposite parity that differ by one unit of principal quantum number are not coupled. An even perturbation cannot couple opposite parity states of the same $k$, $\langle k,+|\sin(\varphi)|k,-\rangle=0$, and does not generate an energy imbalance. Nevertheless, it can couple states that differ by one unit of $k$ and of opposite parity, giving rise to finite matrix elements $\beta_\pm = \langle 0,\pm|\sin(\varphi)|1,\mp\rangle$. All in all a generic $2\pi$-periodic perturbation acts as
\begin{equation}\label{Eq:sin}
    H'=u_o(\alpha_+-\alpha_-\sigma_z)\tau_x+u_e \beta\sigma_x\tau_x
\end{equation}
with $u_{e/o}$ the amplitude of the even/odd perturbation and $\alpha_\pm=(\alpha_0\pm \alpha_1)/2$. In Fig.~\ref{Fig4}a) the microwave spectrum is shown for a pure $\pi$-periodic Hamiltonian $H_0$ with $\lambda=E_M=0$. No transition are allowed in the frequency window $-0.5<\omega/E_C<1.5$. The effect of a $2\pi$-periodic perturbation containing both a $\cos(\varphi)$ and $\sin(\varphi)$ term is shown in Fig.~\ref{Fig4}b), showing the transitions activated by the perturbation. The two branches associated to even and odd total number of fermions in the island are usually seen in the microwave spectrum as a result of an out-of-equilibrium population of the two fermion parity sectors. Finally, in Fig.~\ref{Fig4}c) the presence of finite $E_M=0.2 E_C$ and $\lambda=0.1 E_C$ shows the activation of MQ induced transitions. It follows that the presence of MZMs can be detected, even in presence of weak $2\pi$-periodic perturbations.

{\it Discussion.---}We have shown how a parity-protected superconducting qubit based on an effective $\pi$-periodic JJ can be coupled to a Majorana qubit. The PPSQ is obtained by two highly transparent gate-tunable semiconducting wires providing Andreev states characterized by a high harmonic content. As an alternative to the application of an external magnetic flux $\Phi_0/2$ in the PPSQ loop a $\pi$-junction can be employed, such as the one realized in quantum dot Josephson junctions \cite{bargerbos2022singlet-doublet}. The possibility to tune the semiconductor junctions through side gates enables in principle switching on and off the coupling controlled by $E_M$.  The setup can be generalized to other implementations of a $\pi$-periodic JJ and a three-junction solution in a balanced double ring geometry \cite{ronzani2014balanced} can also be employed, that can further allow  minimization of $2\pi$-periodic perturbations, generally affecting all implementations. The system may suffer from gate errors and dephasing due to the required sensitivity of the spectrum to the offset charge $n_g$ and a mitigation approach can be provided by adiabatic tuning from weak to strong transmon regime through flux-tunable JJs \cite{hassler2011top-transmon}. The setup enables the implementation of a SWAP gate between the two qubits, thus allowing for the use of a Majorana qubit as a memory for storage. 

{\it Acknowledgments.---}The authors acknowledge very useful discussions with M. Carrega, A. Crippa, M. Devoret, V. Fatemi, E. Lee, and V. Manucharyan, and especially thank B. van Heck and R. Aguado for a careful reading of the manuscript and for comments about the evolution of Andreev states in magnetic field. This project has received funding from the European Union’s Horizon 2020 research and innovation programme under the Marie Sklodowska-Curie grant agreement No 841894. J.E.M. was supported by the U.S. Department of Energy, Office of Science through the Quantum Science Center (QSC), a National Quantum Information Science Research Center. N.Y.Y. was supported by the U.S. Department of Energy, Office of Science, through the Quantum Systems Accelerator (QSA), a National Quantum Information Science Research Center and the David and Lucile Packard foundation.

\bibliography{biblio-JJ}{}

\end{document}